\documentclass[12pt]{spieman}  
\usepackage{amsmath,amsfonts,amssymb}
\usepackage{graphicx}
\usepackage{setspace}
\usepackage{tocloft}
\usepackage{amsmath,amsfonts}
\usepackage{algorithmic}
\usepackage[normalem]{ulem}
\usepackage{array}
\usepackage[caption=false,font=normalsize,labelfont=sf,textfont=sf]{subfig}
\usepackage{textcomp}
\usepackage{tabularx}
\usepackage{stfloats}
\usepackage{makecell}
\usepackage{url}
\usepackage{verbatim}
\usepackage{graphicx}
\usepackage{amsmath,amssymb,amsfonts}
\usepackage{algorithmic}
\usepackage{graphicx}
\usepackage{textcomp}
\usepackage{xcolor}
\usepackage{stfloats}
\usepackage{cite}
\usepackage{verbatim}
\usepackage{ragged2e}
\usepackage{caption}
\usepackage{afterpage}
\usepackage{upgreek}
\usepackage{pdfcomment}
\usepackage{graphicx}
\usepackage{textgreek}
\usepackage{amsmath}
\usepackage{siunitx}
\usepackage{stfloats}
\usepackage{makecell}
\usepackage{graphicx}
\usepackage{textcomp}
\usepackage{array} 
\usepackage{longtable}
\usepackage{booktabs}
\usepackage{float}
\usepackage{bm}
\usepackage{stackengine}
\usepackage{balance}
\usepackage{mhchem}
\usepackage{ragged2e}
\usepackage[numbers,sort&compress]{natbib}

\setcitestyle{open={},close={}}
\title{\justifying{Waveguide superlattices with artificial gauge field towards colorless and crosstalkless ultrahigh-density photonic integration}}

\author[a]{Xuelin Zhang}
\author[a]{Jiangbing Du*}
\author[b]{Ke Xu}
\author[a]{and Zuyuan He}
\affil[a]{State Key Laboratory of Advanced Optical Communication Systems and Networks,
Shanghai Jiao Tong University,
800 Dongchuan Road, Shanghai 200240, China}
\affil[b]{Department of Electronic and Information Engineering, Harbin Institute of Technology (Shenzhen), Shenzhen 518055, China} 

\cftpagenumbersoff{figure}
\cftpagenumbersoff{table} 
\begin{document} 
\captionsetup{labelformat=default,labelsep=period}
\maketitle

\begin{abstract}
Dense waveguides are the basic building blocks for photonic integrated circuits (PIC). Due to the rapidly increasing scale of PIC chips, high-density integration of waveguide arrays working with low crosstalk over broadband wavelength range is highly desired. However, the sub-wavelength regime of such structures has not been adequately explored in practice. Herein, we proposed a waveguide superlattice design leveraging the artificial gauge field (AGF) mechanism, corresponding to the quantum analog of field-induced n-photon resonances in semiconductor superlattices.  
This approach experimentally achieves -24 dB crosstalk suppression with an ultra-broad transmission bandwidth over 500 nm for dual polarizations.
The fabricated waveguide superlattices support high-speed signal transmission of 112 Gbit/s with high-fidelity signal-to-noise ratio profiles and bit error rates. 
This design, featuring a silica upper cladding, is compatible with standard metal back end-of-the-line (BEOL) processes. 
Based on such a fundamental structure that can be readily transferred to other platforms, passive and active devices over versatile platforms can be realized with a significantly shrunk on-chip footprint, thus it holds great promise for significant reduction of the power consumption and cost in PICs.

\end{abstract}


{\noindent \footnotesize\textbf{*} Jiangbing Du,  \linkable{dujiangbing@sjtu.edu.cn} }

\begin{spacing}{2}   

\section{Introduction}
\label{sect:intro}  
{
Photonic integrated circuits (PICs) have recently become a mature and powerful technology.
It allows for a variety of applications including optical interconnects,\textsuperscript{\cite{roelkens2010iii}} microwave photonics,\textsuperscript{\cite{marpaung2019integrated}} quantum information,\textsuperscript{\cite{wang2020integrated}}and so forth.
Waveguide arrays are the cornerstones for integrated photonics, which play an ever-increasing role in various functional applications, such as wavelength-division multiplexers,\textsuperscript{\cite{shen2024ultra}} space-division multiplexers,\textsuperscript{\cite{richardson2013space}} mode division 
multiplexers, \textsuperscript{\cite{liu2015densely}} and chip-scale optical interconnections.\textsuperscript{\cite{taubenblatt2011optical}} 
Waveguide arrays possess all the basic characteristics of the photonic crystal structure, including Brillouin zones, allowed bands, and forbidden bands, and therefore support wave dynamics that are equivalent to the dynamics of electron transport in semiconductors.\textsuperscript{\cite{yablonovitch1987inhibited}} 
Periodic modulation of a photonic lattice opens up novel opportunities for diffraction management,\textsuperscript{\cite{szameit2009polychromatic}} manipulating the vortex light,\textsuperscript{\cite{chang2024observation}} and controlling nonlinear interactions of light.\textsuperscript{\cite{matuszewski2006crossover}}

It remains a challenge to significantly improve the comprehensive performance of dense photonic integrated components to fully meet the stringent requirements of large-scale photonic integration.
Substantial research efforts have been made by incorporating superlattices,\textsuperscript{\cite{song2015high,gatdula2019guiding}} inverse design,\textsuperscript{\cite{shen2016increasing}} artificial-gauge-bending  designs,\textsuperscript{\cite{song2022dispersionless,zhou2023artificial,yi2022demonstration}}extreme-skin-depth,\textsuperscript{\cite{jahani2018controlling,mia2020exceptional}} etc.
However, these approaches have mainly focused on silicon-on-insulator (SOI) wafers with air cladding, which have a large refractive index contrast. 
They are not implementable on large-scale optical systems, fundamentally limiting the integration density and scalability of optical chips for practical applications, because the absence of a solid upper cladding destroys the mirror symmetry of unetched waveguides and is incompatible with most metal back-end-of-line (BEOL) processes. 
This incompatibility complicates integration with critical photonic devices such as high-speed modulators and photodetectors.

The silicon nitride (\ce{Si3N4}) photonics,\textsuperscript{\cite{xiang2021high,liu2024parallel,8472140,lu2021efficient,liu2021high}} as a promising photonic integration platform, has facilitated the development of a diverse range of low-loss (\textless 1 dB/m) planar-integrated devices and chip-scale solutions. 
This platform offers unprecedented transparency across a wide wavelength range (400–2350 nm) and enables fabrication through wafer-scale processes. 
Serving as a complementary platform to SOI,\textsuperscript{\cite{lin2022monolithically,li2020experimental}} LNOI,\textsuperscript{\cite{churaev2023heterogeneously}} and III-V photonics,\textsuperscript{\cite{de2020heterogeneous}} \ce{Si3N4} waveguide technology introduces a new era of system-on-chip applications that cannot be achieved solely with other platforms.\textsuperscript{\cite{8472140}}
Nevertheless, the low refractive index contrast and the relatively immature fabrication technology of the \ce{Si3N4} waveguides pose obstacles to directly transplanting the existing dense photonic schemes into the \ce{Si3N4} platform.

In this paper, we propose an alternative approach to realize ultra-broadband (experimentally over 500 nm, theoretically over 1000 nm) nearly zero crosstalk transmission by waveguide superlattices with artificial gauge field using 800-nm-thick \ce{Si3N4} waveguides.
The ultra-dense waveguide arrays with silica upper cladding feature negligible insertion losses and record large bandwidth from 1200 to 1700 nm at a minimum gap (400 nm) with \textless$-24$ dB crosstalk.
Finally, 112 Gbit/s signals encoded on each channel are successfully transmitted along the ultra-compact waveguide arrays with high-fidelity signal-to-noise ratio profiles and bit error rates (BER) below the 7\% HD-FEC threshold.
The strong coupling suppression of the field-induced n-photon resonance, introduced by the curved trajectory of waveguide superlattices, results in an ultra-broadband and dense photonic integrated circuit, which facilitates substantial on-chip footprint reduction and opens up possibilities for high-density heterogeneous integration. 
This work, transferable to other platforms, holds the potential to advance device performance, such as half-wavelength-pitched optical phased array (OPA), high-density energy-efficient modulators, and ultra-dense wavelength-division multiplexers.
}

\vspace{0 cm}
\begin{figure}[!htb]
\begin{center}
\includegraphics[width=0.8\linewidth]{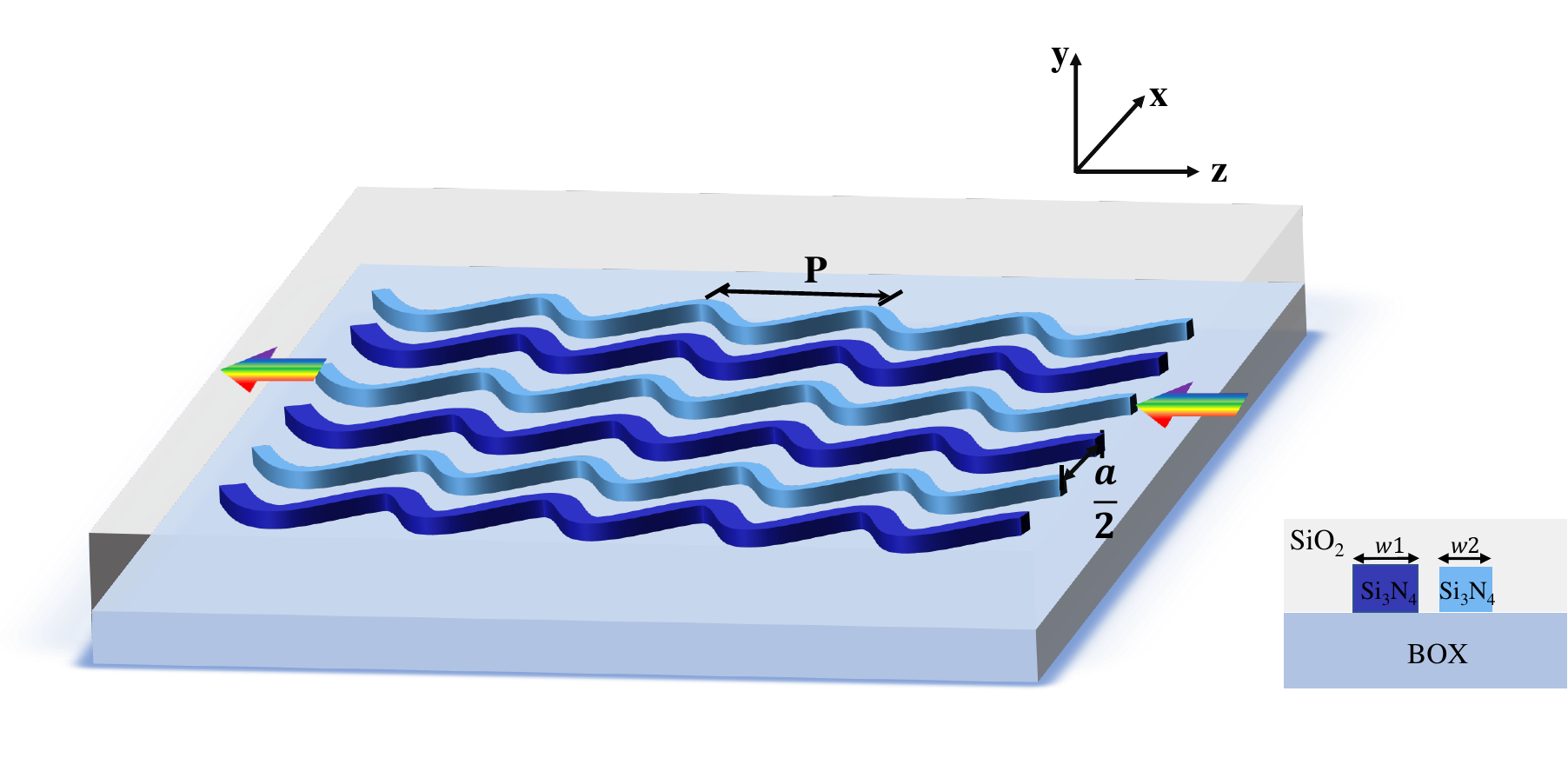}
   \label{fig:1b}
\end{center}
   \caption{Schematic of the waveguide superlattices with artificial gauge field. }
   \vspace{0 cm}
\label{fig:shiyi}
\end{figure}

\begin{figure*}[!t]

\begin{center}
\includegraphics[width=\linewidth]{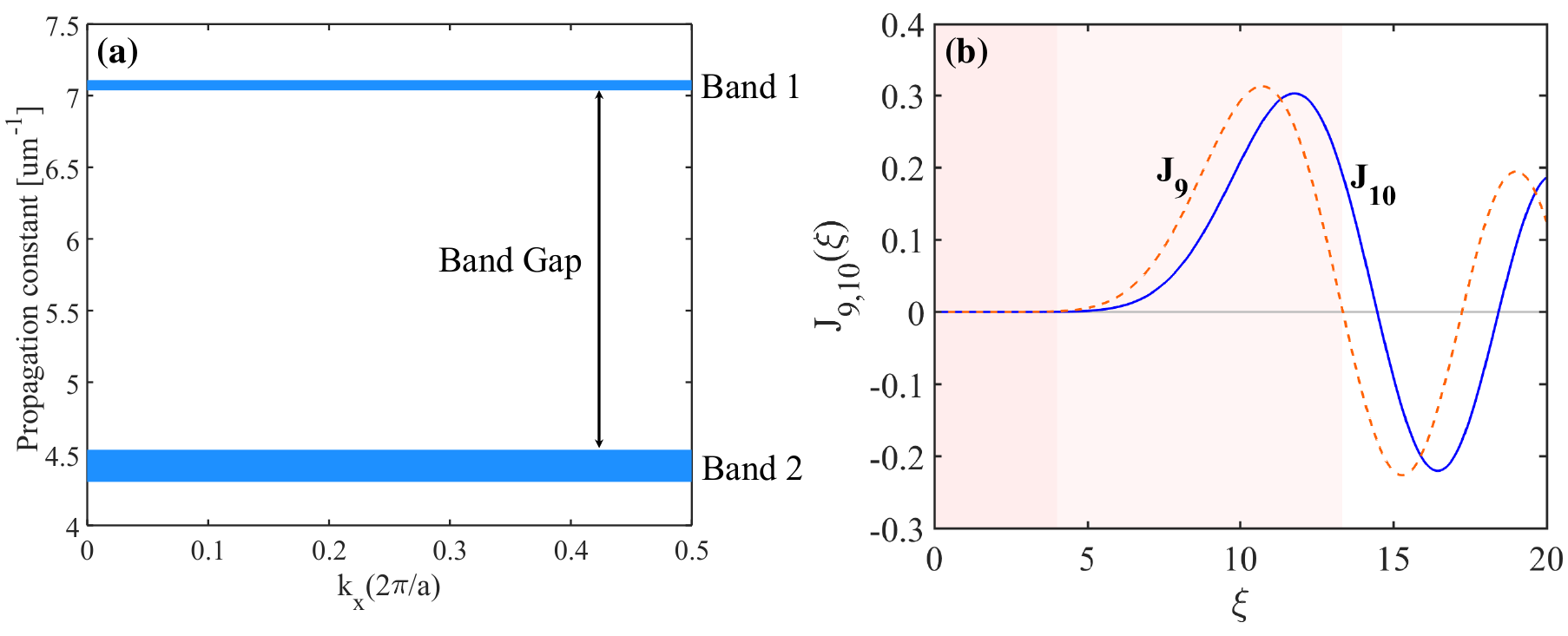}
   \label{fig:1b}
\end{center}
  \caption{(a) Band-gap diagram (first two minibands) of a typical straight waveguide array for a = 2.65 $\upmu$m, $w_{1}$ = 1 $\upmu$m, $w_{2}$ = 0.85 $\upmu$m, $\Delta$n = 0.31, $\lambda$ = 1.55 $\upmu$m, $n_{0}$ = 1.450, where the propagation constant is plotted as a function of the Bloch wave number and the shaded regions represent the bands. (b) Ninth-($J_{9}$) and tenth-order ($J_{10}$) Bessel modulation of the first kind. $J_{10}$ and $J_{9}$ have the same sign in the marked red region.}
  \label{fig:yuanli}
\end{figure*}

\section{Theory and Design Principles}
We first start from a standard 2D model of a binary waveguide array with periodically curved trajectory of frequency along the light-propagation direction, z.
We typically apply a sinusoidal modulation profile 
$x_{0}(z)=Asin(\Omega z)$, where $A$ and $\Omega$ = 2$\pi$/P are the amplitude and period of the trajectory, respectively.
The array consists of a series of equally spaced alternating waveguides of varying width $w_{1}$ and $w_{2}$, separated by $a/2$.
In such a structure shown in
Fig. \ref{fig:shiyi}, light propagation is determined by the overlap between the modes of adjacent waveguides, similar to wave dynamics in discrete lattices.\textsuperscript{\cite{garanovich2012light}} 
Different colors are used to distinguish the waveguides of different widths.
The waveguide superlattices with AGF, as a general one-dimensional periodic optical structure, are analyzed in the framework of the Floquet–Bloch (FB) analysis.\textsuperscript{\cite{longhi2006multiband}}
It predicts that the spectrum of propagation constants for the eigenmodes of the array, known as FB waves, is partitioned into bands with gaps between them where no propagating modes exist.
In the nearest-neighbor tight-binding (NNTB) approximation and assuming that
the lowest Bloch band of the array is excited, the light propagation in a modulated binary array can be fully described based on coupled-mode theory governing the modal amplitudes $\Psi_{n}$ of light waves confined in the individual waveguides number $n$. \textsuperscript{\cite{bohigas1993manifestations}} 
\begin{flalign}\label{eq1}
&\ i \frac{d \Psi_{n}}{d z}+(-1)^{n} \Delta \beta+C\left(\Psi_{n+1}+\Psi_{n-1}\right)=-n F \Psi_{n} &
\end{flalign}
where $z$, 2$\Delta \beta$, and C are the paraxial propagation distance, propagation constant mismatch, and the hopping rate between two neighboring waveguides of the array, respectively, and F is the modulation function related to the waveguide bending.
For straight unmodulated waveguides, where F = 0, a plane wave ansatz $\Psi_{n} \propto$  $\exp(i(\beta z - nkd))$ is inserted into Eq. (1) and the following dispersion relation for the two minibands is obtained as follows:
\begin{flalign}\label{eq2}
&\ \beta=\pm\sqrt{\Delta\beta^2+4C^2\cos^2(ka)}&
\end{flalign}
where $\beta$, $k$, and $a$ are the longitudinal, transverse propagation constants, and the period of the binary array, respectively.
Fig. \ref{fig:yuanli}\textcolor{blue}{(a)} shows the calculated band-gap diagram of the straight array, establishing the relationship between the propagation constant and the Bloch wave number.
Here, $n_{0}$ represents the substrate refractive index, and $\Delta$n is the effective refractive index difference between silicon nitride and silicon dioxide.
It is important to note that the two minibands are separated by a gap $E_{g} = 2\Delta \beta$.
A nontrivial AGF can be introduced, considering the sinusoidal modulation function which follows an arbitrary periodic function $x(z)=x(z+P)$.
\begin{flalign}\label{eq3}
&\
F=\frac{4\pi^2\omega A}{P^2}\sin\left(\frac{2\pi z}P\right)
&
\end{flalign}
where $\omega$ is the normalized optical frequency. 

The quantum mechanical description of the light propagation dynamics in the proposed waveguide array, as given by Eq. (1), when combined with the modulation described by Eq. (3), i.e., a sinusoidal driving field coupled with a periodic width modulation, is equivalent to the motion of a charged particle within a two-site crystalline potential, subject to the action of an external field of frequency $\Omega$ = 2$\pi$/P. 
The condition $E_{g} = n\Omega$ when the ratio $n$ is an integer, where $E_{g} = 2\Delta \beta$ represents the width of the gap in the band-gap diagram, corresponds to field-induced n-photon resonances between the two minibands. 
Light dynamics in the modulated waveguide array can be captured by the effective equations under this condition of n-photon resonance,
\begin{flalign}\label{eq4}
&\
i\frac{d\tilde{\Psi}_n}{dz}+C_n(\tilde{\Psi}_{n+1}+\tilde{\Psi}_{n-1})=0
&
\end{flalign}
where the effective coupling coefficient can be derived. It is $C_{n_eff} = CJ_{n}(\pi \omega A/P)$ for even resonances (n even), and 
$C_{n_eff} = (-1)^{n}CJ_{n}(\pi\omega A/P)$ for odd resonances (n odd). 
Here $J_{n}$ is the Bessel function of the first kind of the order n.
Note that the output field pattern strongly relies on the normalized modulation depth $\xi=2\pi^2An_0(\lambda)a/P\lambda$, which signifies the existence of bending-induced resonances between narrow and wide waveguides in the modulated waveguide array.
When the parameter $\xi$ takes the roots of the Bessel function $J_{n}$, an effective suppression of waveguide coupling is attained.
This observation shows great potential for realizing zero crosstalk in dense integration.

Then, we focus on the dispersion of coupling to investigate the wavelength dependence of waveguide superlattices with a well-defined AGF.
\begin{flalign}\label{eq4}
\begin{split}
\frac{\partial c_{\mathrm{eff}}(\lambda)}{\partial\lambda}=(-1)^{n}
\left\{J_n(\xi)\left\{\frac{\partial c(\lambda)}{\partial\lambda}+
c(\lambda)\frac{n}{\lambda}\left[1-\frac{\partial n_{0}(\lambda)/\partial\lambda}{n_{0}(\lambda)/\lambda}\right]\right\} \right. \\ 
\left. - J_{n-1}
(\xi)c(\lambda)\frac{\xi}{\lambda}\left[1-\frac{\partial n_{0}(\lambda)/\partial\lambda}{n_{0}(\lambda)/\lambda}\right]\right\}
\end{split}&
\end{flalign}

Intuitively, the effective coupling dispersion can be fully suppressed as long as
$J_{n}$ and $J_{n-1}$ 
share the same signs.
The unusual modulation-induced dispersion properties of this engineered waveguide array enable the elimination of wavelength dependence in crosstalk reduction.
It is worth noting that the insensitivity to wavelength also indicates robust performance against changes in key structural parameters such as waveguide spacing, the amplitude of
the trajectory, and the widths of the waveguides. 
We calculate the derivative of effective coupling $c_{eff}$ concerning $a$ as an example to clarify the mechanism of robustness:
\begin{flalign}\label{eq6}
&\
\frac{\partial c_{\mathrm{eff}}(a)}{\partial(a)}=(-1)^{n}
\left\{J_{n}(\xi)\{\frac{\partial c(a)}{\partial(a)}-\frac{n}{d}c(a)\} +
J_{n-1}(\xi)c(a)\frac{2\pi^2An_0(\lambda)}{P\lambda}\right\}
&
\end{flalign}
Equation \ref{eq6} verifies that this modulation can also mitigate the structural sensitivity to waveguide separation, provided that $J_{n}$ and $J_{n-1}$ have the same sign.

Based on these analyses, we design the AGF-enabled densely packed waveguide superlattices on a standard 800-nm-thick \ce{Si3N4} platform with a silica upper cladding. 
We choose the waveguide widths of 1300, 1150, 1000, 850 nm, and 1400 nm constrained by the single-mode condition at 1550 nm.
The waveguide superlattices follow a sinusoidal curve, with modulation period P fixed at 25 $\upmu$m, which satisfies the field-induced n-photon resonances of $n=10$.
Fig. \ref{fig:yuanli}\textcolor{blue}{(b)} shows the  $J_{9}$ and $J_{10}$ functions, and they indeed have the same sign in the marked region.
It is found that $J_{9}$ and $J_{10}$ are positive integers arbitrarily close to 0 when the normalized modulation depth $\xi$ is in the dark red area.
Near the zero effective coupling condition, i.e., $J_{10} \approx 0$, and $J_{9}$ is also almost zero, which indicates the intrinsic waveguide coupling and coupling dispersion are fully suppressed.
Therefore, this design provides flexibility for engineering coupling coefficient and dispersion, thereby achieving broadband and robust zero-coupling performance for densely packed waveguides.

\section{Characterizations of the  Waveguide Superlattices with AGF}

\begin{figure*}[!htb]
\begin{center}
\includegraphics[width=\linewidth]{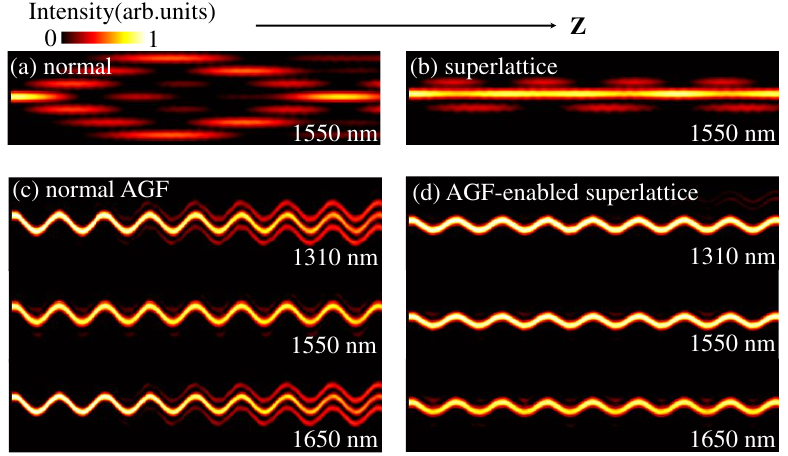}
   \label{fig:1b}
\end{center}
   \caption{Simulated normalized field evolution in (a) normal waveguides, (b) the ordinary superlattice, (c) the AGF waveguide array, and (d) AGF enabled superlattice, where gap = 400 nm, $L$ = 200 $\upmu$m.}
\label{fig:field}
\end{figure*}
\vspace{0 cm}

We simulate light evolution in the modulated and normal (i.e., without modulation) waveguide array in Fig. \ref{fig:field}, respectively, to make a clear comparison. 
The straight waveguide superlattice has the same width parameters as the AGF-enabled superlattice.
The normal AGF waveguide array has been optimized to the best level and the width of the single-mode waveguides is set to 1000 nm, the Width of a standard single-mode waveguide.
Due to the lower refractive index, the widths of single-mode \ce{Si3N4} waveguides are larger than that of SOI waveguides at the same wavelength, thus their pitches are also larger.
For the AGF-enabled superlattice in Fig. \ref{fig:field}\textcolor{blue}{(d)}, the light remains localized in the input waveguide with almost no coupling to other waveguides at all three wavelengths (1310 nm, 1550 nm, and 1650 nm), owing to the bending-induced resonances between narrow and wide waveguides.
The nearly isolated guidance indicates zero crosstalk between adjacent waveguides.
It can be seen that our modulated waveguides exhibit high through transmission across a very broad bandwidth from the O-band to extend beyond the C-band.
In contrast, if either form of the modulation (the width modulation or sinusoidal trajectory modulation) is removed, the light will couple with other waveguides, and the crosstalk between them will drastically increase.
The crosstalk in normal AGF waveguides in Fig. \ref{fig:field}\textcolor{blue}{(c)} could be suppressed to some extent at 1550 nm, while the useable bandwidth is significantly limited (see Fig. S2  in the \textcolor{blue}{Supplementary Material}).
The conventional one in Fig.\ref{fig:field}\textcolor{blue}{(a)} exhibits a discrete diffraction phenomenon and drastic changes in coupling length attributed to the strong dispersion.
Although the conventional superlattice and AGF waveguide array has successfully demonstrated low crosstalk on the standard silicon-on-insulator platform with air cladding at C-band, the deleterious effects of manufacturing defects are exacerbated by the high refractive index contrast between silicon and air. 
This exposure becomes a barrier to heterogeneous integration, significantly limiting the scalability and integration of the photonic chip. 
Besides, the use of silica upper cladding could reduce phase errors, which is 
greatly needed in phase-sensitive applications such as optical-phased
arrays and high-speed modulators.\textsuperscript{\cite{wang2024chip}} 
The above-mentioned crosstalk suppression mechanisms in Fig.\ref{fig:field}\textcolor{blue}{(a)-\ref{fig:field}(c)} are not strong enough when meeting the platform with a low refractive index contrast and usually work only for a particular band, showing the weaknesses of their schemes.

\begin{figure*}[!h]
\begin{center}
\includegraphics[width=\linewidth]{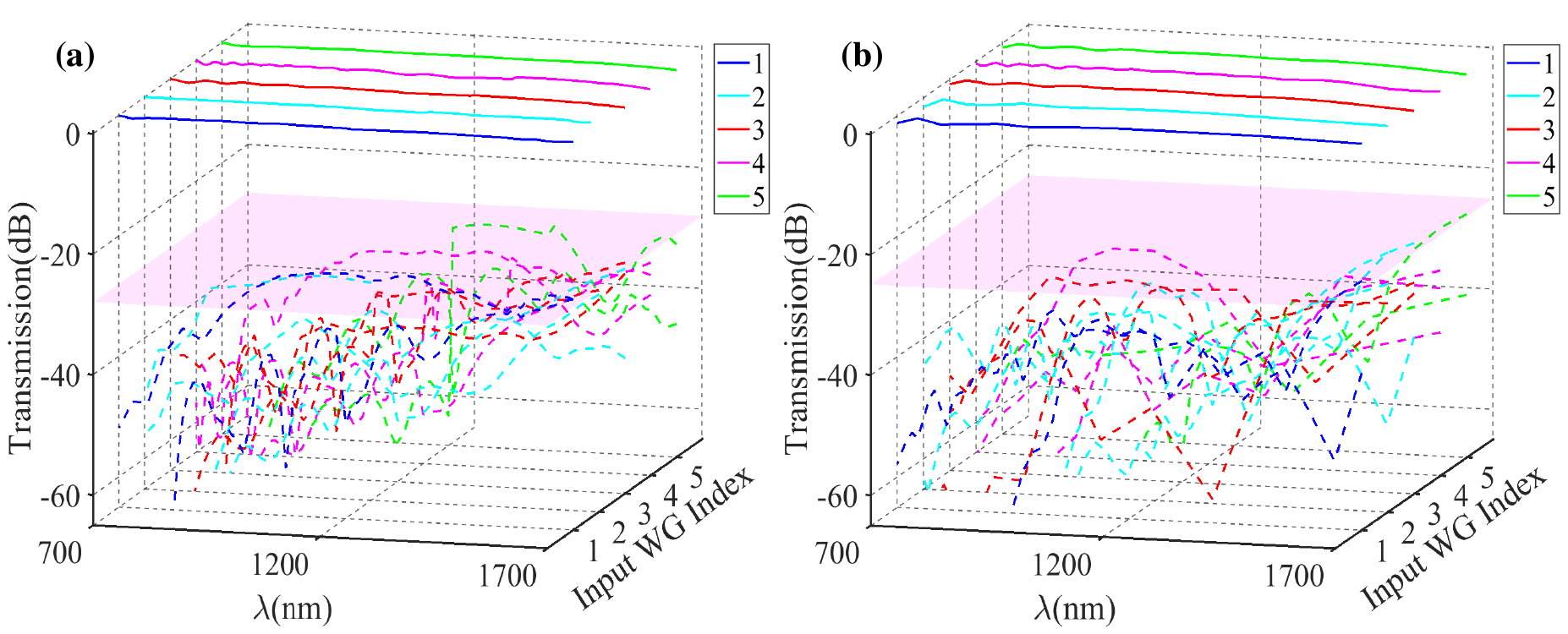}
   \label{fig:1b}
\end{center}
  \caption{Simulated crosstalk in the designed waveguide array when
   (a) fundamental TE mode and (b) fundamental TM mode are respectively launched (gap = 400 nm, $\lambda$ =700-1700 nm). The pink planes correspond to $- 28$ dB and $- 25$ dB, respectively.}
  \label{fig:simu}
\end{figure*}

The simulated transmission of the proposed dense waveguide array is depicted in Fig. \ref{fig:simu}.
Most of the previous designs work for one polarization only (TE or TM polarization), the calculated effective indices of the fundamental TE and TM modes of 800-nm thick \ce{Si3N4} waveguides are close enough to support dual polarizations (see Fig. S1  in the \textcolor{blue}{Supplementary Material}).
Although a slight increase in crosstalk can be observed at longer wavelengths, the crosstalk for the fundamental TE and TM modes remains less than 25 dB in the spectrum of 700–1700 nm, opening up possibilities for colorless and crosstalk-free ultrahigh-density photonic integration.


\begin{figure*}[!htb]
\begin{center}
\includegraphics[width=1\linewidth]{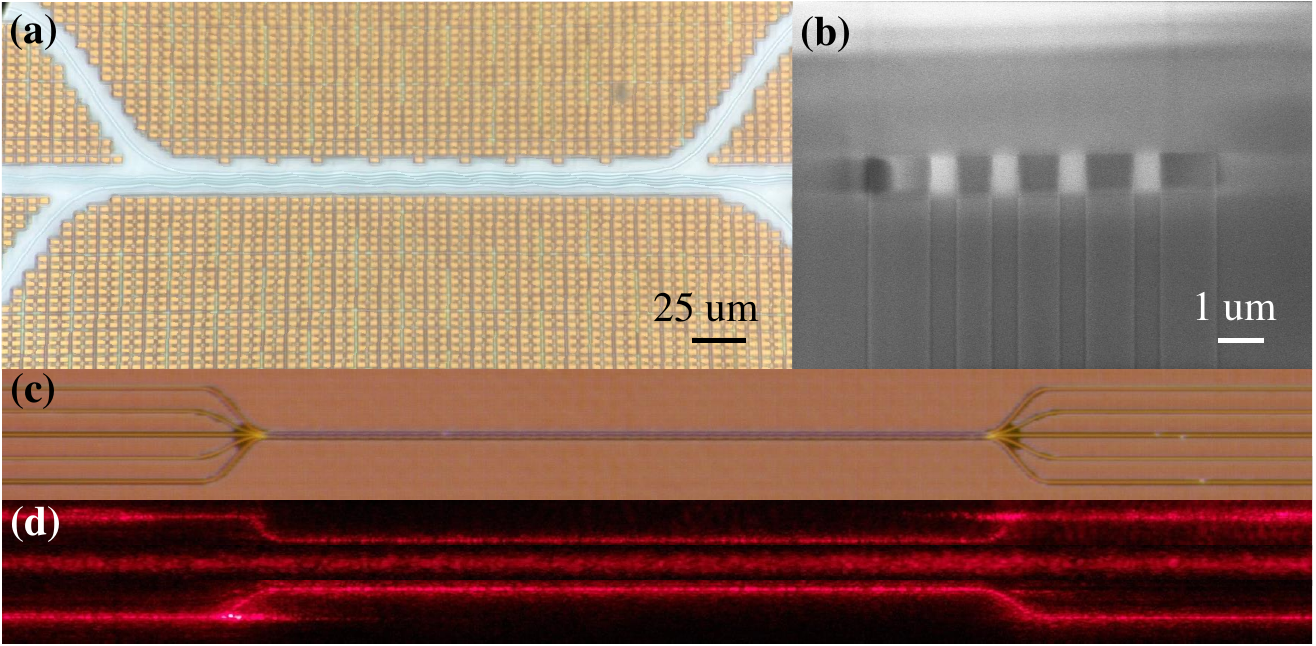}
   \label{fig:1b}
\end{center}
   \caption{(a) The top-view microscope image of the fabricated device on the \ce{Si3N4} platform. (gap = 400 nm, $L$ = 200 $\upmu$m). (b) SEM image of the cross-section of a waveguide (gap = 400 nm, $L$ = 200 $\upmu$m). (c) The top-view microscope image of the fabricated device on the \ce{Si3N4} platform. (gap = 500 nm, $L$ = 1 mm). (d) Experimentally recorded light trajectory in modulated samples for 850 nm.}
\label{fig:shiwu}
\end{figure*}
\vspace{0 cm}

\begin{figure*}[!htb]
\begin{center}
\includegraphics[width=1\linewidth]{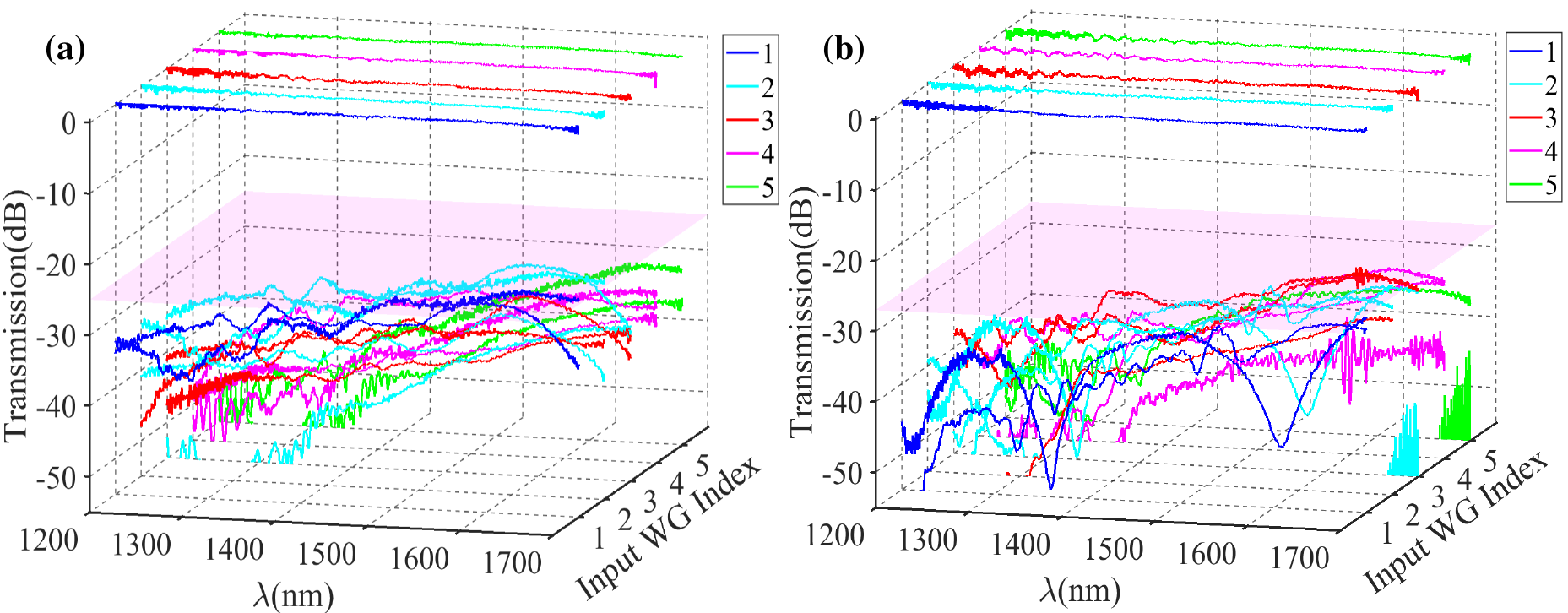}
   \label{fig:1b}
\end{center}
   \caption{(a) Measured transmission spectra of the AGF-enabled superlattice (gap = 400 nm). (b) Measured transmission spectra of the AGF-enabled superlattice (gap = 500 nm). The pink planes correspond to $- 24$ dB and $- 28$ dB, respectively.}
\label{fig:ct}
\end{figure*}

In experiments, we characterize the proposed waveguide arrays with different pitches, i.e., gap = 400, 500, and 600 nm.
The sample length is 200 $\upmu$m for a gap of 400 nm, and more than 1 mm for gaps larger than 400 nm. 
Fig. \ref{fig:shiwu}\textcolor{blue}{(a)} and \ref{fig:shiwu}\textcolor{blue}{(b)} show the microscope image and the scanning electron microscope (SEM) image of a cross-section of the fabricated device, respectively, where the width and sinusoidal trajectory variation can be observed. 
The edge couplers are used for light input and output, which could achieve broadband and dual-polarization operation.
After the light is coupled into the chip (see Fig. S6 in the \textcolor{blue}{Supplementary Material}), it propagates along the single-mode waveguide and enters the dense waveguide array horizontally.
Fig.\ref{fig:shiwu}\textcolor{blue}{(d)} displays the experimentally captured optical propagations for the modulated samples at 850 nm.
Measured normalized transmission spectra of the AGF-enabled superlattice for gap = 400 nm and 500 nm are displayed in 
Fig. \ref{fig:ct}\textcolor{blue}{(a)} and Fig. \ref{fig:ct}\textcolor{blue}{(b)}, respectitively.
Crosstalk suppression occurred over an ultra-wide wavelength range from 1200 to 1700 nm, where the crosstalk is less than $-24$ dB in this entire 500 nm bandwidth.
It can be seen that the crosstalk was decreasing as the waveguide array became sparse.
Although the crosstalk suppression was strong enough for most practical applications, it could be further reduced by increasing the bending radius of interfaces between the separated waveguides and the dense waveguide array.
Moreover, the measured bandwidth of the device was limited by the device instead of the proposed AGF-enabled superlattice.

\vspace{0 cm}

\begin{figure*}[!htb]
\begin{center}
\includegraphics[width=1\linewidth]{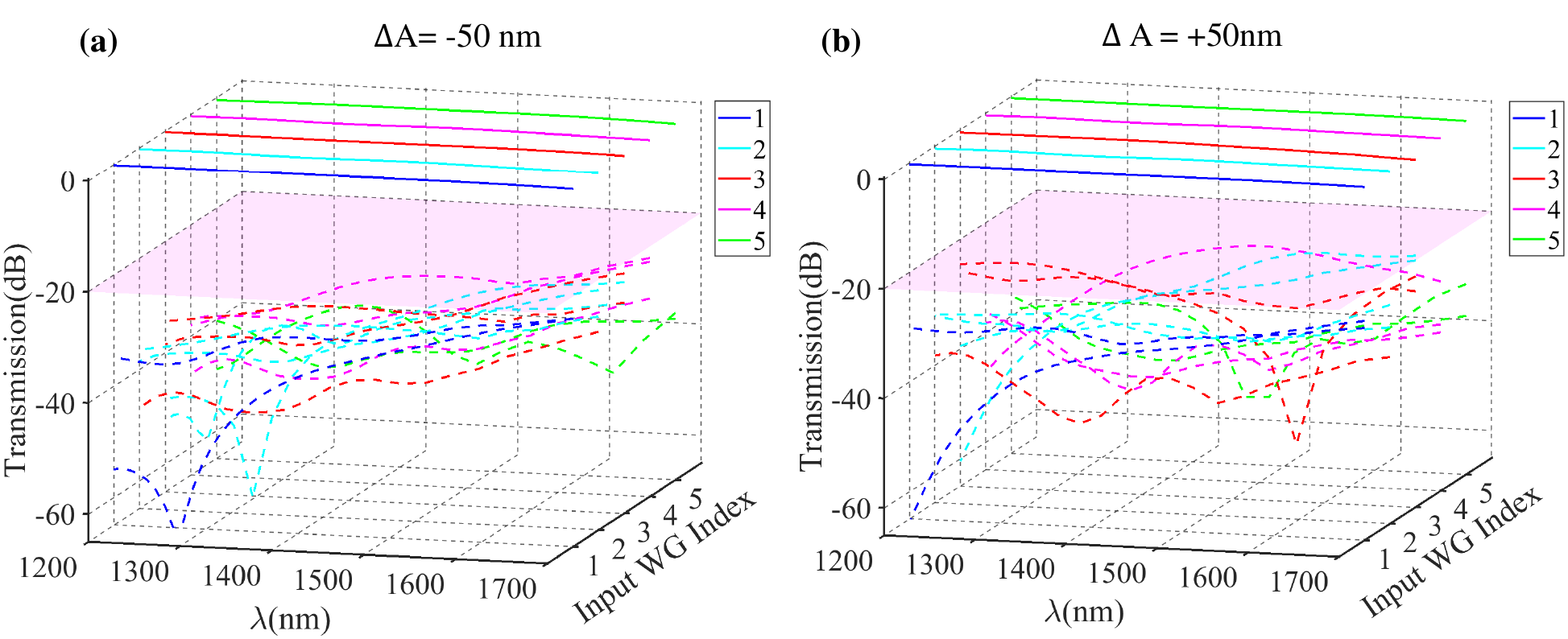}
   \label{fig:1b}
\end{center}
   \caption{Simulated crosstalk for AGF-enabled superlattice with (a) $\Delta$ A = -50 nm and (b) $\Delta$ A = +50 nm. The pink planes correspond to $- 20$ dB.}
\label{fig:A}
\end{figure*}

\begin{figure*}[!htb]
\begin{center}
\includegraphics[width=1\linewidth]{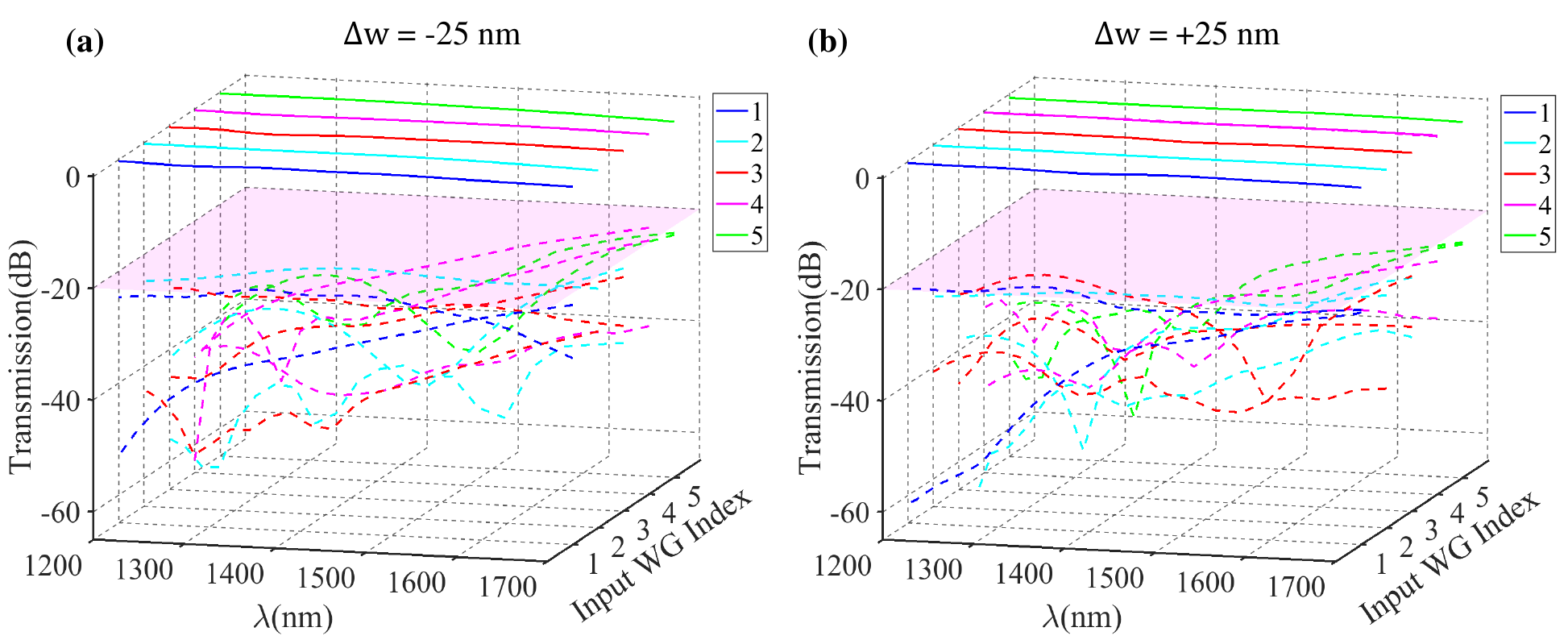}
   \label{fig:1b}
\end{center}
   \caption{Simulated crosstalk for AGF-enabled superlattice that widths of all waveguides (a) increase by 25 nm and (b) decrease by 25 nm. The pink planes correspond to $- 20$ dB.}
\label{fig:w}
\end{figure*}

To further clarify the mechanism of robustness, we simulate the transmission spectra of the AGF-enabled superlattice with the variations of A and w at 50 nm and 25 nm in Fig. \ref{fig:A} and Fig. \ref{fig:w}, respectively. 
The results show that despite introducing significant structural parameter changes, they still exhibit excellent crosstalk suppression and negligible insertion loss across the entire frequency spectrum for all channels.
The deliberately modified devices can keep crosstalk below $-20$ dB with a bandwidth exceeding 500 nm, demonstrating the robustness of our scheme.
This robustness ensures a large tolerance to dimensional uncertainties in the fabrication process, allowing scalability to large-scale circuits.

\vspace{0 cm}

\vspace{0 cm}

\begin{figure*}[!htb]
\begin{center}
\includegraphics[width=\linewidth]{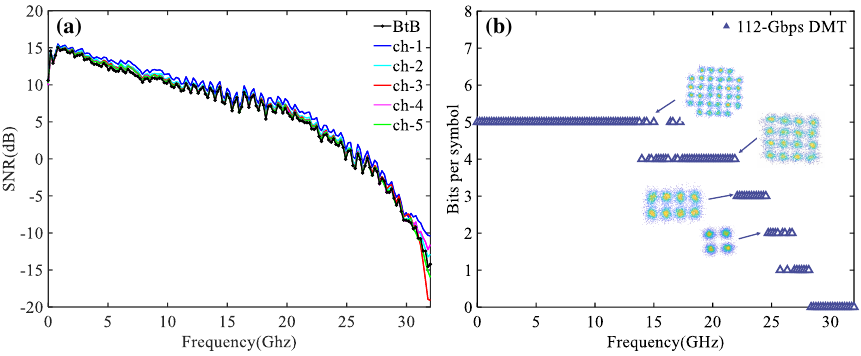}
   \label{fig:1b}
\end{center}
   \caption{(a) Measured signal-to-noise-ratio (SNR) profiles of the system after transmission through the waveguide array. (b) The bit allocation of 112-Gbit/s DMT signals. Inset: constellations of different subcarriers.}
\label{fig:snr}
\end{figure*}
\vspace{0 cm}




Alongside the measurements of loss and crosstalk, we also verified the ability of the proposed waveguide array to perform high-speed signal routing (see Fig. S3 in the \textcolor{blue}{Supplementary Material} for more details on the experimental setup and digital signal processing techniques).
The signal-to-noise-ratio (SNR) profiles 
and the bit allocation were obtained and plotted in Fig. \ref{fig:snr}\textcolor{blue}{(a)} and \ref{fig:snr}\textcolor{blue}{(b)}.
The SNR profiles of the entire transmission link include the AWG, the electrical driver, the waveguide array, the modulator, the photodetector (PD), and the digital sampling oscilloscope (DSO). 
The transmission through each channel of the waveguide array achieves nearly identical SNR performance as the BtB case, indicating excellent signal quality.
The roughly unchanged end-to-end SNR profiles, combined with extensively suppressed SNR performance in the adjacent waveguides (see Fig. S4 in the \textcolor{blue}{Supplementary Material}), suggest that the transmitted signals in the excited waveguide do not affect the neighboring waveguides, which is indicative of good isolation between waveguides and effective crosstalk suppression in the spectrum.
In optical communication systems, it is desirable for each waveguide (channel) to operate independently without interfering with each other.
SNR measurements can verify the signal isolation performance and spectral integrity of the proposed waveguide array, ensuring that each channel in a multi-channel system can reliably transmit signals.
The SNR performance was used to generate the DMT signal through the bit allocation algorithm, as shown in Fig. \ref{fig:snr}\textcolor{blue}{(b)}.
The maximum bit allocation is 5, corresponding to 32-QAM.
The constellations in the inset of \ref{fig:snr}\textcolor{blue}{(b)} indicate good signal quality for high-order modulation formats (see Fig. S5 in the \textcolor{blue}{Supplementary Material}).

\begin{figure*}[!htb]
\begin{center}
\includegraphics[width=\linewidth]{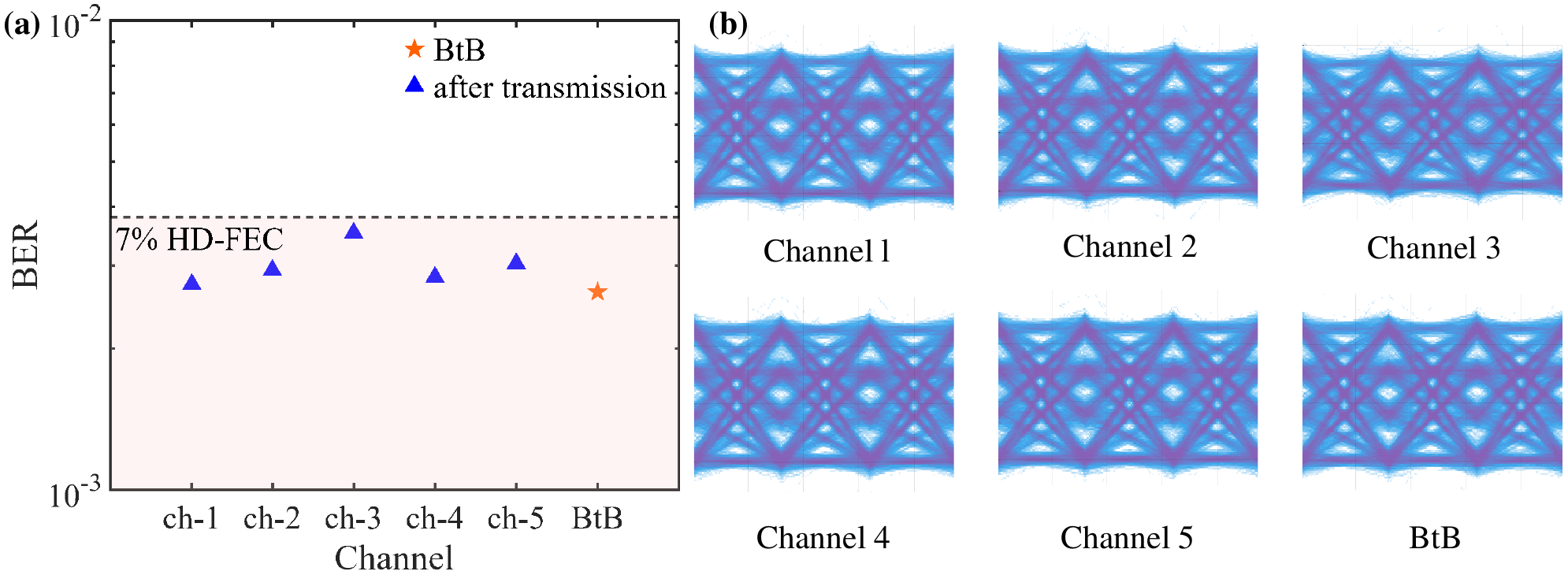}
   \label{fig:1b}
\end{center}
   \caption{(a) The BER of 112 Gbit/s PAM4 after transmission through modulated samples. (b) Received eye diagrams of PAM4 signals.}
\label{fig:pam}
\end{figure*}
\vspace{0 cm}



Besides, we conducted the high-speed 112 Gbit/s PAM4 signal transmission experiment through the AGF-enabled waveguide superlattice.
Fig. \ref{fig:pam}\textcolor{blue}{(a)} and 
Fig. \ref{fig:pam}\textcolor{blue}{(b)} depict the bit error rate (BER) and eye diagrams after back-to-back (BtB) and five-channel transmission, respectively.
The BER after transmission is below the 7\% HD-FEC threshold ($3.8 \times 10^{-3}$).
Thus the ultra-compact signal routing of high-speed signals has been successfully achieved.
The eye diagram in BtB transmission is clearly open, and we obtain almost the same eye diagrams after transmitting through the ultra-dense waveguides.
The differences in the transmission performance, due to insertion losses, crosstalk, or even environmental instability, are roughly imperceptible compared with the BtB case. 
All the waveguide channels
present almost identical BER and SNR performances relative to the BtB case, demonstrating the proposed low-crosstalk and ultra-compact waveguide array is capable of high-speed on-chip transmission.

\section{Discussion}
The waveguides are perhaps the most fundamental building blocks in integrated photonics. However, coupling between closely packed waveguides has been a long-term challenge due to significant crosstalk at small pitches.
The high-density waveguide arrays demonstrated in this study can effectively address this issue.
The engineering capability of the exceptional coupling is presented both theoretically and experimentally, and it provides a new flexible toolbox for densely packed power-efficient photonic components.
For example, shrinking the spacing of an optical phased array to half wavelength could lead to superior beam characteristics with a 180-degree field of view.\textsuperscript{\cite{leng2021waveguide}}
Dense waveguide arrays enable the development of highly efficient thermo-optic phase shifters and compact switches.\textsuperscript{\cite{liu2022thermo}}
Besides, the penalty-free transmission of high-speed signals and the support of advanced modulation formats through the proposed array can stimulate new directions in broadband, compact, and high-speed optical modulators.

It is worth noting that the dimensional control requirements of the proposed approach, unlike the nanohole
metamaterials,\textsuperscript{\cite{yi2024chip}} are well within the capabilities of state-of-the-art foundries, making it amenable to large-scale production.\textsuperscript{\cite{smith2023sin}}
The millimeter-level lengths of the waveguide array in this work are sufficient for most applications such as ultra-dense wavelength-division multiplexers,\textsuperscript{\cite{richardson2013space}}on-chip spectrometers, \textsuperscript{\cite{li2022advances}} and optical phased arrays.\textsuperscript{\cite{liu2022silicon,fukui2021non}}
The specially designed superlattice structure, combined with artificial gauge field, enables precise control over the light propagation characteristics, achieving colorless and crosstalk-free optical transmission and laying the foundation for the miniaturization and cost-effective of advanced photonic systems.

\section{Conclusion}
In summary, we have experimentally demonstrated an ultra-broadband, low-crosstalk dual-polariz\-ation \ce{Si3N4} waveguide superlattice leveraging artificial gauge field.
The physical principle is attributed to the bending-induced resonance between the narrow and wide waveguides, corresponding to the field-induced n-photon resonance between the two mini bands, which results in a broadband zero-coupling effect.
This coupling mechanism achieves a working bandwidth exceeding 500 nm for the crosstalk below $-24$ dB.
Given its compatibility with BEOL processes, this well-designed waveguide array already serves as a robust solution for dense waveguide integration.
When combined with the potential for transferability to other platforms, it will be an attractive strategy to significantly improve the waveguide density limit and performance capabilities of various active and passive photonic devices and systems, such as half-wavelength spacing OPAs, high-density on-chip optical interconnects, and energy-efficient modulators.
Our proposed scheme also offers a versatile arrangement to study a range of fascinating scientific phenomena, including Rabi oscillations.

\subsection* {Acknowledgments}
This work was supported by the National Key Research and Development Program of China (2023YFB2905502); the National Natural Science Foundation of China (62122047, 61935011). 

\bibliography{ref}
\bibliographystyle{spiejour}









\end{spacing}
\end{document}